\documentclass[aps,prd,showpacs,superscriptaddress,amssymb,amsmath,
nofootinbib,twocolumn]{revtex4}

\usepackage{graphicx}
\usepackage{verbatim}
\usepackage{amsthm}
\usepackage{amsfonts}
\usepackage{amscd}

\usepackage{array}

\begin{document}


\title{Electromagnetic Leptogenesis}

\author{Nicole~F.~Bell}
\affiliation{School of Physics, 
The University of Melbourne, Victoria 3010, Australia}

\author{Boris~J.~Kayser}
\affiliation{Theoretical Physics Department, 
Fermilab, PO Box 500, Batavia, IL 60510-0500, USA}

\author{Sandy~S.~C.~Law}
\affiliation{School of Physics, 
The University of Melbourne, Victoria 3010, Australia}

\date{\today}

\begin{abstract}
  We present a new leptogenesis scenario, where the lepton asymmetry is
  generated by $CP$ violating decays of heavy electroweak singlet
  neutrinos via electromagnetic dipole moment couplings to the
  ordinary light neutrinos. Akin to the usual scenario where the
  decays are mediated through Yukawa interactions, we have shown, by
  explicit calculations, that the desired asymmetry can be produced
  through the interference of the corresponding tree-level and
  one-loop decay amplitudes involving the effective dipole moment
  operators.  We also find that the relationship of the leptogenesis
  scale to the light neutrino masses is similar to that for the
  standard Yukawa-mediated mechanism.
\end{abstract}

\pacs{13.15.+g, 14.60.St, 13.40.Em   \hspace{4cm} FERMILAB-Pub-08/186-T}




\maketitle


\section{Introduction}\label{sec:intro}

Baryogenesis via thermal leptogenesis~\cite{Fukugita:1986hr} provides
an elegant explanation of the cosmic baryon asymmetry~\cite{wmap}. The
conventional setup involves minimally extending the Standard Model
(SM) by adding three heavy right-handed (RH) Majorana neutrinos, which
are electroweak singlets, and allowing them to interact with ordinary
left-handed (LH) lepton doublets via complex Yukawa couplings. As a
result, when the heavy neutrinos decay out-of-equilibrium in a $CP$
violating way, a lepton asymmetry is generated in the early
Universe. This asymmetry is then partially converted by sphaleron
processes to the baryon asymmetry we detect today.

An attractive aspect of this scenario is that it naturally allows the
implementation of the see-saw mechanism~\cite{seesaw_list} which gives
the light neutrinos a tiny but nonzero mass.~\footnote{Note that
  similar results can be instead achieved with the addition of heavy
  Higgs triplets~\cite{Ma:1998dx}, or with other
  mechanisms~\cite{zeemodel}.} Consequently, a strong link between some
  neutrino properties and successful asymmetry generation can be
  established. For instance, to simultaneously obtain a sufficiently
  large lepton asymmetry and the correct light neutrino masses, the
  heavy neutrino masses must be larger than $~10^9$~GeV in most viable
  leptogenesis scenarios~\cite{Ibarra}.

In this paper, we will consider a scenario where leptogenesis is
mediated not by the standard Yukawa couplings, but instead by
electromagnetic dipole moment couplings.  One motivation for
introducing such couplings is to explore whether new sources of $CP$
violation can lead to a significant lepton asymmetry.  A natural
question we may ask is whether the introduction of $CP$ violating
dipole moment couplings will allow leptogenesis to occur at a lower
scale, closer to experimentally accessible energies.  However, we find
this not to be possible due to the connection between the dipole
moment couplings and the neutrino mass.

The general form of a dipole moment coupling of the light neutrinos,
$\nu$, to the heavy neutrinos, $N$, is given by $\overline{\nu} (\mu+
i\, d\gamma_5)\sigma^{\alpha\beta}N F_{\alpha\beta}$, where $\mu$ and
$d$ are the magnetic and electric transition moments, respectively.
These dimension-5 effective operators may be assumed to be generated
by some new physics beyond the electroweak scale.  While we do not
speculate on the nature of this new physics, the inclusion of these
operators permits a new leptogenesis mechanism.  Radiative decays of
the heavy neutrinos, $N \rightarrow \nu +\gamma$, can now produce the
required lepton asymmetry in the early Universe, provided that the
complex electromagnetic dipole moment couplings violate $CP$.

Below, we outline the relevant properties of the electromagnetic
dipole moment (EMDM) couplings, and discuss the necessary requirements
for a decay process to manifestly violate $CP$.  We shall then
explicitly calculate the decay rates and $CP$ asymmetry for
electromagnetic leptogenesis, and compare with the standard
Yukawa-mediated leptogenesis scenario.


\section{Electromagnetic coupling between light and heavy neutrinos}\label{sec:emag}

We extend the particle content of the minimal SM by adding heavy RH
neutrinos, $N_R$, which are assumed to have large Majorana masses.
Since we are interested in leptogenesis energy scales above the
electroweak phase transition, we will take the usual light neutrinos
to be massless LH states.  The most general electromagnetic dipole
moment coupling of the heavy RH neutrinos to the light LH neutrinos is
then given by
\begin{equation}\label{equ:EMDM_lag1}
{\cal H}_\textrm{EM} = g_{ij} \overline{\nu}_{Li}\, 
\sigma^{\alpha\beta} N_{Rj} F_{\alpha\beta} + \textrm{h.c.},
\end{equation}
where $g$ is a complex (dimensionful) matrix, and $i,j,$ are flavor
indices.  Note that there is only one distinct electromagnetic dipole
moment coupling when expressed in term of LH and RH chiral fields
(rather than distinct magnetic and electric dipole moment terms) since
$\gamma_5 P_{\textrm L,R} \propto P_{\textrm L,R}$.


\section{$CP$ violation in decays}

If the dipole coupling of Eq.(\ref{equ:EMDM_lag1}) had
less-constrained chiral structure, then for a given $\nu_i$ and $N_j$
there could be independent magnetic and electric transition moments,
$\mu_{ij}$ and $d_{ij}$. It might be thought that in this situation, a
tree-level interference between the amplitude induced by $\mu_{ij}$
and that induced by $d_{ij}$ could lead to a $CP$-violating difference
between the rates for $N_j \rightarrow \nu_i + \gamma$ and its $CP$
conjugate. However, there can never be a difference between the rates
for $CP$-conjugate decay modes until one goes beyond first order in
the underlying Hamiltonian. This fact is well known, but it is
interesting to see that it can be proved very simply by using $CPT$
invariance. Consider, in the rest frame of the parent particle $Q$,
the decay $Q \rightarrow a_1 + a_2 +\cdots$. If $CPT$ invariance
holds, the amplitude for this decay obeys the constraint
\begin{align}
 &|\langle a_1(\vec p_1, \lambda_1)\, a_2(\vec p_2, \lambda_2)\cdots |\mathcal{T}|Q(\widehat m) \rangle|^2
  \nonumber\\
  &\qquad 
  =
   |\langle \overline{a}_1(\vec p_1, -\lambda_1)\, 
   \overline{a}_2(\vec p_2, -\lambda_2)\cdots |\mathcal{T}^\dagger|\overline{Q}(-\widehat m) \rangle|^2\;.
   \label{equ:x1}
\end{align}
Here, $\vec p_i$ and $\lambda_i$ are, respectively, the momentum and
helicity of daughter particle $a_i$, $\widehat m$ is the $z$-axis
projection of the spin of $Q$, and $\mathcal{T}$ is the transition
operator for the decay. If $S$ is the $S$-matrix operator,
$\mathcal{T}=i(S-I)$. To first order in the Hamiltonian $\mathcal{H}$
for the system, $\mathcal{T}=\mathcal{H}$, so that, to this order,
$\mathcal{T}^\dagger = \mathcal{T}$. From the latter relation and
Eq.(\ref{equ:x1}), it follows that, after summing over the final
helicities and integrating over the outgoing momenta,
\begin{equation}
 \Gamma\left[\overline{Q} \rightarrow \overline{a}_1 +\overline{a}_2 +\cdots \right]=
  \Gamma\left[Q \rightarrow a_1 +a_2+\cdots\right]\;.
\end{equation}
This equality must hold to first order in $\mathcal{H}$ regardless of
whether $\mathcal{H}$ contains numerous terms and $CP$-violating
coupling constants.

In the special case of a two-body decay, $Q \rightarrow a_1 + a_2$, we
have $\vec p_1 = -\vec p_2 \equiv \vec p$. For this case, let us rotate the
system of particles on the right-hand side of Eq.(\ref{equ:x1}) by
$180^\circ$ about the axis perpendicular to the $z$-axis and to $\vec
p$. Eq.(\ref{equ:x1}) then states that, to first order in
$\mathcal{H}$ (so that $\mathcal{T} = \mathcal{T}^\dagger$),
\begin{align}
 &|\langle a_1(\vec p, \lambda_1)\, a_2(-\vec p, \lambda_2)|\mathcal{T}|Q(\widehat m) \rangle|^2
 \nonumber\\
&\quad\qquad
= 
  |\langle \overline{a}_1(-\vec p, -\lambda_1)\, 
   \overline{a}_2(\vec p, -\lambda_2)|\mathcal{T}|\overline{Q}(\widehat m) \rangle|^2\;.
\end{align}
The processes whose amplitudes appear on the two sides of this
constraint are the $CP$-mirror images of each other. Thus, in two-body
decays, to first order in $\mathcal{H}$, the rates for
$CP$-mirror-image decay processes must be equal even before one sums
over final helicities and integrates over outgoing momenta.

We conclude that $CP$-violating rate differences between
$CP$-conjugate electromagnetic decays of heavy neutrinos can only
appear once amplitudes involving loops are included.


\section{A toy model}\label{sec:toy_model}

In this section we illustrate, by means of a toy model, the viability
of generating a lepton asymmetry through EMDM interactions between
ordinary light neutrinos and the postulated heavy Majorana neutrinos
in the early universe.  
In order to illustrate the physics as transparently as possible, we
begin by considering a simplistic model which is not invariant under
the SM gauge symmetry, $SU(2)_L \times U(1)_Y$, but is instead
invariant only under the electromagnetic symmetry $U(1)_Q$.  We will
generalize to a realistic model in which invariance under the SM gauge
group is enforced in section~\ref{subsec:3body} below.

We assume the EMDM couplings are generated by new physics at an energy
scale $\Lambda>M$, where $M$ denotes a heavy Majorana neutrino mass.
We work with an effective theory that is valid below the scale
$\Lambda$, obtained after integrating out all new heavy degrees of
freedom.  The lowest dimension EMDM operator of interest in such a
scenario is given by:
\begin{align}
\mathcal{L}_\text{EM}^\text{5D}=
   -\frac{1}{\Lambda}e^{-i\varphi_k/2}   
   \lambda_{jk}\, 
   \overline{\nu}_{Lj}\,  
    \sigma^{\alpha\beta} P_R N_{k} F_{\alpha\beta}
   +\text{h.c.} \,, \label{equ:lag_5d}
\end{align}
where $j=e,\mu,\tau$ and $k = 1,2,3$.  The electromagnetic field
strength tensor is $F_{\alpha\beta} =
\partial_{\alpha}A_{\beta}-\partial_{\beta}A_{\alpha}$, with
$A_\alpha$ being the photon field.  The light neutrinos are denoted by
$\nu_{j}$, while $N_{k}$ is the heavy neutrino field in the mass
eigenbasis which satisfies the Majorana condition: $N_{k} =
e^{i\varphi_k} N_{k}^c$, for some arbitrary phase $\varphi_k$. We have
defined $\lambda$ as a dimensionless $3\times 3$ matrix of complex
coupling constants, while $\Lambda$ is the cut off scale of our
effective theory. For convenience, we have factored out
$e^{-i\varphi_k/2}$ for each $N_k$ in Eq.(\ref{equ:lag_5d}) so that
the Majorana phases will not appear explicitly in any of our final
expressions.

\begin{figure}[tb]
\begin{center}
\includegraphics[width=0.5\columnwidth]{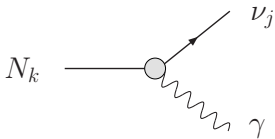}
\caption{The tree-level diagram for the decay of $N_k$ via the EMDM
  interaction of Eq.~(\ref{equ:lag_5d}).}
\label{fig:tree5d}
\end{center}
\end{figure}

\begin{figure}[tb]
\centering
\includegraphics[height=0.29\columnwidth]{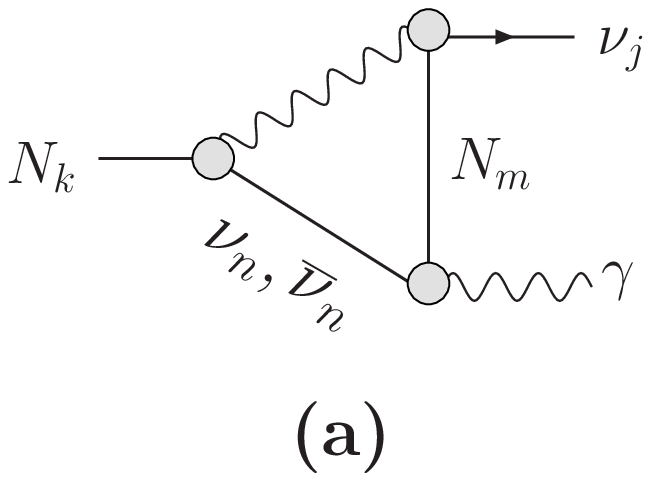}
\hspace{-3mm}
\includegraphics[height=0.29\columnwidth]{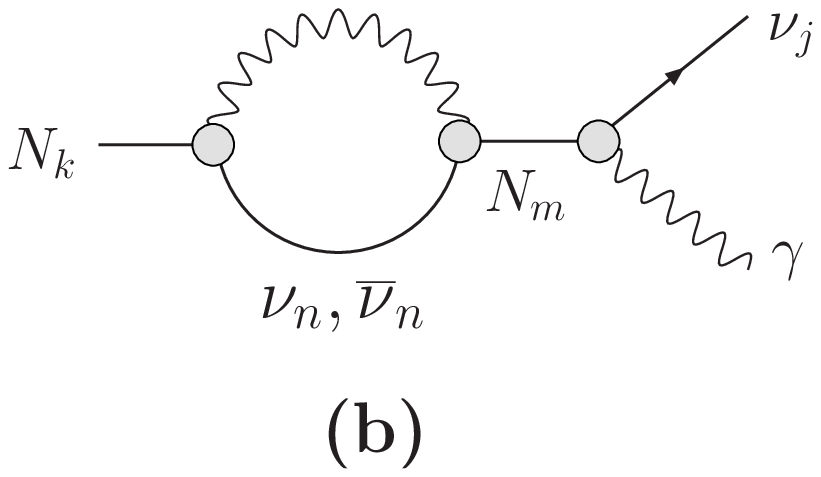}
\caption{(a) Vertex and (b) self-energy diagrams which contribute to
  the $CP$ asymmetry of $N_k$ decay via the interaction of
  Eq.(\ref{equ:lag_5d}). Note that since weak isospin is violated in
  this model, both $\nu_n$ and $\overline{\nu}_n$ are allowed in the
  loop of (a).}
\label{fig:loop5d}
\end{figure}

To ascertain whether leptogenesis is possible, the key quantity of
interest is the $CP$ asymmetry in the decays of $N_k$:
\begin{equation}
 \varepsilon_{k, j}^{(5)} = \frac{
 \Gamma_{(N_k\rightarrow \nu_j\,\gamma)}
 -\Gamma_{(N_k\rightarrow \overline{\nu}_j\,\gamma)}}
 {\Gamma_{(N_k\rightarrow \nu\,\gamma)}
 +\Gamma_{(N_k\rightarrow \overline{\nu}\,\gamma)}}\;,
 \label{equ:emag5}
\end{equation}
where $\Gamma_{(N_k\rightarrow \nu\,\gamma)} \equiv \sum_j
\Gamma_{(N_k\rightarrow \nu_j\,\gamma)}$. 
The lowest order contribution to the decay rate, shown in
Fig.~\ref{fig:tree5d}, is given by
\begin{equation}
\Gamma_{(N_k\rightarrow \nu\,\gamma)}
 =\Gamma_{(N_k\rightarrow \overline{\nu}\,\gamma)}
 =\frac{(\lambda^{\dagger} \lambda)_{kk}}{4\pi}\,
\frac{M_k^3}{\Lambda^2}\;.
\label{equ:tree5d}
\end{equation}
%
The leading contribution to the $CP$ asymmetry, $\varepsilon_{k,
  j}^{(5)}$, comes from the interference of the tree-level process of
Fig.~\ref{fig:tree5d} with the 1-loop diagrams with on-shell
intermediate states depicted in Fig.~\ref{fig:loop5d}.  As with
standard (Yukawa interaction) leptogenesis, the $CP$ asymmetry
receives two contributions: the self-energy and vertex correction. We
have calculated these explicitly and obtain
\begin{align}
 \varepsilon_{\text{self-}k,j}^{(5)} &=
  \frac{-(M_k/\Lambda)^2}{2\pi(\lambda^\dagger \lambda)_{kk}} 
  \sum_{m\neq k}
  \text{Im} \bigg[\lambda_{jk}^* \lambda_{jm}  
\nonumber\\
  & \times \left\{
  (\lambda^\dagger \lambda)_{km}\;
   \frac{\sqrt{z}}{1-z}
   +(\lambda^\dagger \lambda)_{mk}\;
   \frac{1}{1-z}
   \right\}   
   \bigg] ,
   \label{equ:self5d}
\end{align}
for the self-energy contribution and
\begin{align}
 \varepsilon_{\text{vert-}k,j}^{(5)} &=
    \frac{-(M_k/\Lambda)^2}{2\pi(\lambda^\dagger \lambda)_{kk}}
  \sum_{m\neq k}
  \text{Im} \left[\lambda_{jk}^* \lambda_{jm} 
  (\lambda^\dagger \lambda)_{km}\;
  f(z)
  \right]\;,\label{equ:vert5d}\\
\nonumber
\text{ with }
 &f(z) 
  = \sqrt{z}\left[1+2z\left(1-(z+1)\ln \left[\frac{z+1}{z}\right]\right)\right]\;,
\end{align}
for the vertex piece, where $z\equiv M_m^2/M_k^2$.  Note that in
Eq.(\ref{equ:self5d}) and Eq.(\ref{equ:vert5d}), we have not yet
summed over the final lepton flavor $j$.

The expressions given in Eq.(\ref{equ:self5d}) and
Eq.(\ref{equ:vert5d}) are very akin to those in standard leptogenesis
\cite{Covi:1996wh_Liu:1993tg}.  The first and second terms of
Eq.(\ref{equ:self5d}) correspond to the interference terms involving
$\overline{\nu}_n$ and $\nu_n$ as the intermediate state in
Fig.~\ref{fig:loop5d}(b) respectively. It should be emphasized that
upon summing over $j$ (ie. ignoring flavor effects
\cite{Nardi:2006fx}), the second term vanishes. Furthermore, explicit
calculations have shown that in Fig.~\ref{fig:loop5d}(a), the
contribution from the $\nu_n$ intermediate state actually evaluates to
zero, and hence there is no term proportional to $(\lambda^\dagger
\lambda)_{mk}$ in Eq.(\ref{equ:vert5d}).

The total $CP$ asymmetry from the decays of $N_k$'s into light
neutrinos and photons in this model is simply given by
$\varepsilon_{k, j}^{(5)} = \varepsilon_{\text{self-}k,j}^{(5)}+
\varepsilon_{\text{vert-}k,j}^{(5)}$.  This asymmetry would be nonzero
as long as there are phases in the coupling matrix $\lambda$ which
cannot be removed by redefinitions of the neutrino fields. As
$\lambda$ is an arbitrary complex matrix, it is not hard to see that
one cannot eliminate all the relevant phases to render both
Eq.(\ref{equ:self5d}) and Eq.(\ref{equ:vert5d}) zero.  Hence, this
type of EMDM interaction between light and heavy neutrinos will in
general generate a lepton asymmetry in the early Universe.
Before discussing the magnitude of this asymmetry, we will first
generalize this scenario to one in which the EMDM couplings respect
the SM gauge symmetries.


\section{A more realistic extension}\label{subsec:3body}

While the simplistic model in section~\ref{sec:toy_model} can
demonstrate the viability of lepton generation through EMDM operators,
it is nonetheless unrealistic as it is incompatible with the SM. We
now overcome this by considering only EMDM type operators that respect
the SM gauge group.  Again, we construct an effective theory by taking
the usual minimally extended SM Lagrangian with three generations of
heavy Majorana neutrinos, and augmenting it with EMDM operators.  The
most economical of such operators involving only (the minimally
extended) SM fields are of dimension six~\cite{Dirac}, and the
interaction Lagrangian of interest is
\begin{align}
  \mathcal{L}_\text{EM} &= 
- \frac{1}{\Lambda^2} e^{-i\varphi_k/2} \,\overline{\ell}_j
  \left[ \lambda'_{jk} 
\phi\,\sigma^{\alpha\beta}
    B_{\alpha\beta} 
  \right.\nonumber\\
&   \left.\qquad\qquad\qquad
+  \widetilde{\lambda}'_{jk}
\tau_i \phi\,\sigma^{\alpha\beta} W_{\alpha\beta}^i
  \right]P_R N_k
  +\text{h.c.} \,, 
\label{equ:lag_6d}
\end{align}
where the $\tau_i$ are the $SU(2)_L$ generators, $\ell_{j} = (\nu_{j},
e^{-}_{j})^T$ is the lepton doublet, and $\phi =
(\phi^{0*}, -\phi^-)^T$ is the SM Higgs doublet.  The field strength
tensors of $U(1)_Y$ and $SU(2)_L$ are given by $B_{\alpha\beta} =
\partial_{\alpha}B_{\beta}-\partial_{\beta}B_{\alpha}$ and
$W^i_{\alpha\beta} =
\partial_{\alpha}W^i_{\beta}-\partial_{\beta}W^i_{\alpha} - g
\epsilon_{imn} W^m_{\alpha}W^n_{\beta}$, respectively, where $g'$ and
$g$ are the corresponding coupling constants.
Again, $\Lambda$ is the high energy cut off of our effective theory,
while the matrices of dimensionless coupling constants,
$\lambda'$ and $\widetilde{\lambda}'$, are in general complex.

The higher dimension (non-renormalizable) operators of
Eq.(\ref{equ:lag_6d}) are assumed to be generated at the energy scale
$\Lambda$, beyond the electroweak scale.  The presence of these
operators would imply the existence of some new physics at a high
energy.  After $SU(2)_L\otimes U(1)_Y$ breaking, these operators
will give rise to electromagnetic dipole transition moments of $N$ and
$\nu$. However, for the purposes of leptogenesis, we are interested
here in the regime above the electroweak symmetry breaking scale.

\begin{figure}[t]
\centering
\includegraphics[height=0.3\columnwidth]{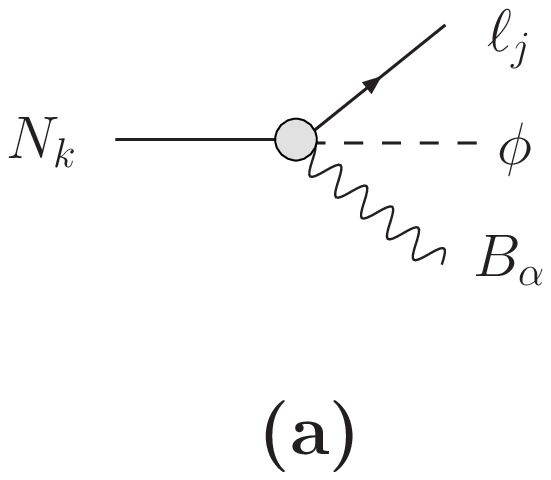}
\hspace{-3mm}
\includegraphics[height=0.3\columnwidth]{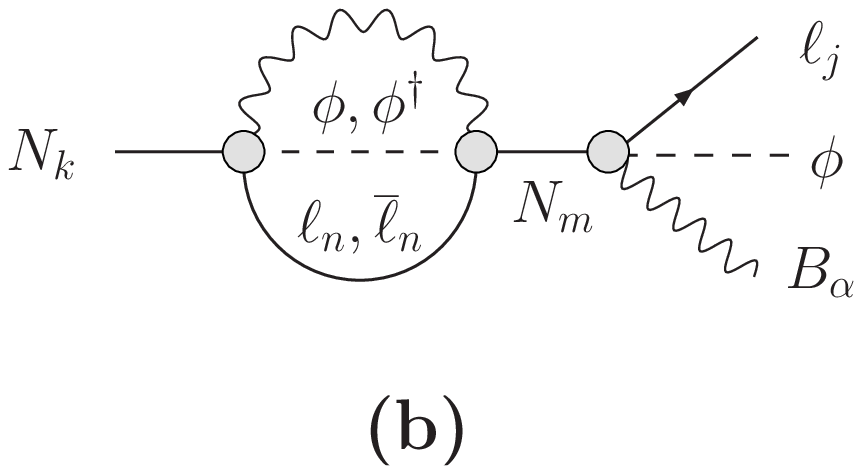}
\caption{(a) The tree-level diagram for the the 3-body decay:
  $N_k\rightarrow \ell_j \phi B_\alpha$ induced by the first term in
  Eq.(\ref{equ:lag_6d}).  (b) The corresponding self-energy
  diagram.}
\label{fig:6d}
\end{figure}

The decay of $N$ will now produce 3-body final states, namely
$N_k\rightarrow \ell_j\phi W^i_\alpha$ and $N_k\rightarrow \ell_j\phi
B_\alpha$, as shown in Fig.~\ref{fig:6d}(a).  Likewise, the
self-energy and vertex corrections now become two-loop processes, an
example of which is shown in Fig.~\ref{fig:6d}(b).  As before, a lepton
asymmetry will be generated through the interference of the tree-level
amplitude with the on-shell part of the vertex and self-energy
amplitudes.

\begin{figure}[t]
\centering
\includegraphics[width=0.45\columnwidth]{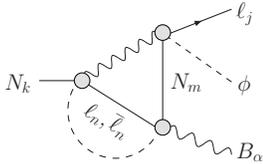}
\caption{An example of a vertex correction to the 3-body decay process.}
\label{fig:6dv}
\end{figure}

To demonstrate that the EMDM interactions in this model can indeed
generate a lepton asymmetry, we have explicitly calculated the
tree-level decay rate, $\Gamma_{(N_k\rightarrow \ell_j\phi
  B_\alpha)}$, and self-energy contribution to the $CP$ asymmetry,
$\varepsilon_{\text{self-}k,j}^{(6)}$, for the couplings which
involve the hypercharge boson $B$ (see Fig.~\ref{fig:6d}).  The
$SU(2)$ gauge boson will make a contribution to the decay rate and $CP$
asymmetry of similar magnitude.  For simplicity, we have not
explicitly calculated the vertex corrections, an example of which is
shown in Fig.~\ref{fig:6dv}.  The vertex diagrams can again have on-shell
intermediate states and, barring accidental cancellations, would
contribute to the lepton asymmetry generated.

For the diagrams shown in Fig.~\ref{fig:6d}, our explicit calculations
of $\Gamma$ and $\varepsilon$ lead to similar forms to those found
earlier:\footnote{It should be noted that these 3-body decay processes are inherently different from those discussed in \cite{Hambye:2001eu} for these are 2-loop rather than 1-loop diagrams.}
\begin{align}
 \Gamma_{(N_k\rightarrow \ell_j\phi B_\alpha)}
  &= 
  \left(\frac{M_k}{8\pi\Lambda}\right)^2 \Gamma_{(N_k\rightarrow \ell_j\gamma)} \;,\label{equ:6dtree}\\
\text{and}
 \qquad\qquad 
  |\varepsilon_{\text{self-}k,j}^{(6)}| &= \left(\frac{M_k}{8\pi\Lambda}\right)^2 
  |\varepsilon_{\text{self-}k,j}^{(5)}|\;,\label{equ:6dself} 
\end{align}
where we must replace $\lambda \rightarrow \lambda'$ in the RHS of
Eq.(\ref{equ:6dtree}) and Eq.(\ref{equ:6dself}).
It is thus possible to generate a non-zero $CP$ asymmetry via the EMDM
type interactions of Eq.(\ref{equ:lag_6d}), provided
$\lambda'$ contains complex phases.

\section{Neutrino mass}

We now address the connection between the dipole moment operators and
the neutrino mass.  The new physics that gives rise to the effective
operators in Eq.~(\ref{equ:lag_6d}) might also be expected to give
rise to neutrino mass terms.  It is well known that via a careful
choice for the new physics, one can obtain large neutrino dipole
moments without correspondingly large mass
terms~\cite{Barr,Voloshin,Georgi,Grimus,Mohapatra}.
However, even if neutrino masses are absent or suppressed at lowest
order, radiative corrections involving the dipole moment operators
generically induce neutrino mass terms~\cite{Dirac,Davidson,Majorana}.

\begin{figure}[tb]
\begin{center}
\includegraphics[width=0.6\columnwidth]{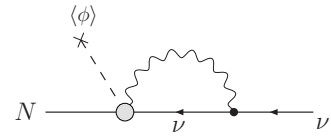}
\caption{
  Contribution to the neutrino Dirac mass, $m_D$, induced by the
  electromagnetic dipole moment operator.}
\label{fig:diracnumass}
\end{center}
\end{figure}

\begin{figure}[tb]
\begin{center}
\includegraphics[width=0.6\columnwidth]{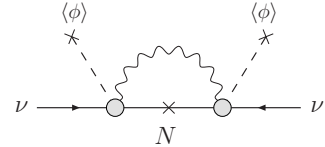}
\caption{
  Contribution to the light neutrino mass, $m^B_\nu$, induced by the
  electromagnetic dipole moment operator.}
\label{fig:lightnumass}
\end{center}
\end{figure}

As in Ref.~\cite{Dirac}, the operators in Eq.~(\ref{equ:lag_6d}) will
lead to a Dirac mass term for the neutrinos via the diagram shown in
Fig.~\ref{fig:diracnumass}.  There is no model-independent way of
calculating the exact size of this neutrino mass contribution, since
the exact relationship between the dipole moments and the mass
requires a UV completion of the theory (i.e. it depends on the nature
of the physics at scale $\Lambda$).  However, we may estimate the size
of the contribution to the neutrino mass using na\"ive dimensional
analysis.  The neutrino Dirac mass arising from
Fig.~\ref{fig:diracnumass} is thus estimated to be
\begin{equation}
m_D \sim \frac{\lambda'}{\Lambda^2} \frac{g'}{16 \pi^2}
\langle \phi \rangle \Lambda^2,
\end{equation}
where the $\Lambda^2$ in the numerator arises from the cut off of the
loop integral, and $g'$ is the gauge coupling constant.  This Dirac
mass term will lead to a contribution to the light neutrino mass via
the see-saw mechanism of
\begin{equation}
m^A_\nu
=  m_D^T  M^{-1} m_D
\sim \lambda'^T M^{-1} \lambda' \langle \phi \rangle^2  
\left( \frac{g'}{16\pi^2} \right)^2.
\end{equation}
However, there will also be a direct contribution to Majorana mass of
the light neutrinos via the diagram in Fig.~\ref{fig:lightnumass},
which we estimate as
\begin{equation}
m^B_\nu \sim 
\lambda'^T M \lambda' 
\frac{\langle \phi \rangle^2 }{\Lambda^2} \frac{1}{16\pi^2}.
\end{equation}

In the standard Yukawa-mediated leptogenesis scenario, the light
neutrino masses are linked with the leptogenesis parameters, since the
Yukawa coupling constants that control the decay rate of the $N$ also
appear in the Dirac mass terms and thus (via the seesaw mechanism) in
the $\nu$ masses.  In order for these Yukawa coupling constants to
give rise to the correct values for both $\varepsilon$ and $m_\nu$,
the heavy neutrinos are required to have masses $M\agt 10^{9}$
GeV~\cite{Ibarra}~\footnote{This condition may be alleviated in
  degenerate leptogenesis scenarios, in which the $CP$ asymmetry can
  be enhanced and thus the $N$ mass lowered through a resonance in the
  self-energy contribution when $M_k \simeq M_m$~\cite{resonant}.}.
The presence of the neutrino mass term arising from the diagrams of
Fig.~\ref{fig:diracnumass} and Fig.~\ref{fig:lightnumass} signifies that
a similar connection will hold for the couplings $\lambda'$ that
control electromagnetic leptogenesis.  Hence, we will also need a
relatively high $N$ mass scale in order to produce a sufficiently large
asymmetry together with sufficiently small $\nu$ masses.

\begin{table}[h]
\begin{center}
\begin{tabular*}{0.95\columnwidth}{@{\extracolsep{\fill}}cc|c}
\hline\hline
Yukawa && Electromagnetic \\ [10pt]
\hline\hline 
$\Gamma_1=\dfrac{1}{8\pi} (h^\dagger h)_{11} M_1$ & &
$\Gamma_1=\dfrac{1}{2\pi} (\lambda'^\dagger \lambda')_{11} M_1
\left(\dfrac{M_1^2}{8\pi\Lambda^2}\right)^2$
\\ [10pt]
\hline
$\varepsilon \sim \dfrac{1}{8\pi}
\dfrac{\textrm{Im}(h^\dagger h)^2_{1m}}{(h^\dagger h)_{11}}\dfrac{M_1}{M_m}$ 
& &
$\varepsilon \sim \dfrac{1}{2\pi}
\dfrac{\textrm{Im}(\lambda'^\dagger \lambda')^2_{1m}}{(\lambda'^\dagger \lambda')_{11}}\dfrac{M_1}{M_m} \left(\dfrac{M_1^2}{8\pi\Lambda^2}\right)^2 $ \\ [10pt]
\hline
$m_\nu \sim h^T M^{-1} h \langle \phi \rangle^2$ & &
$m^A_\nu \sim 
\lambda'^T M^{-1} \lambda' \langle \phi \rangle^2  
\left( \dfrac{g'}{16\pi^2} \right)^2$
\\ [10pt]
& &
$m^B_\nu \sim \dfrac{\lambda'^T M \lambda'}{\Lambda^2} 
\langle \phi \rangle^2 \dfrac{1}{16\pi^2}\quad$ 
\\ [10pt]
\hline\hline  
\end{tabular*}
\end{center}
\caption{Comparison of key quantities in standard and electromagnetic
  leptogenesis, where $h$ denotes Yukawa coupling constants.  We have
  assumed there is at least a mild hierarchy in the masses of the
  heavy neutrinos, such that the asymmetry is predominantly generated
  from the decay of the lightest state, $N_1$.}
\label{table:compare}
\end{table}


\section{Discussion}

If we assume there is at least a mild hierarchy in the masses of the
the heavy neutrinos, the asymmetry will be predominantly generated
from the decay of the lightest state, $N_1$.  In
Table~\ref{table:compare}, we compare the decay rate, $CP$ asymmetry,
and light neutrino masses for electromagnetic leptogenesis with the
corresponding expressions for the standard (Yukawa) scenario.
The coupling constants and the $N_1$ mass enter into the expressions
for $\Gamma_1$, $\varepsilon$, and $m_\nu$ in essentially the same way
for the two scenarios.  Modulo the additional suppression factors in
the RHS of the table, we see that the region of viable parameter space
for electromagnetic leptogenesis must be similar to that for the
standard Yukawa mechanism.
It is also clear that although we require $\Lambda > M$ in order for
our effective operator approach to be valid, we do not want $\Lambda \gg
M$, as it would suppress both $\Gamma$ and $\varepsilon$.  We
therefore require a moderate hierarchy between $M$ and $\Lambda$ in
order to obtain an appropriately sized $CP$ asymmetry.

In general, both the Yukawa and EMDM interactions will contribute to
the decay rate, $CP$ asymmetry, and neutrino mass.  Depending on the
relative size of $h$ and $\lambda'$, either one mechanism will dominate
or there will be an interplay between the two.
In what follows, we suppose that the Yukawa couplings are negligible.
For simplicity, we also ignore the flavour structure of the matrices
$\lambda'$, $m_\nu$, and $M$, and make the crude assumption that all
elements are of similar magnitude.
If we then take 
\begin{equation}
\Lambda \sim 10 M_2 \sim 20 M_1, \quad \textrm{and} \quad \lambda' \sim 35,
\label{eq:Mvalues}
\end{equation}
we obtain an asymmetry of $\varepsilon \sim 10^{-6}$, while
smaller values of $\lambda'$ would lead to a correspondingly smaller
asymmetry according to $\varepsilon \propto (\lambda')^2$.
We define the decay parameter as
\begin{equation}
K \equiv \Gamma_1/H|_{T=M_1},
\end{equation} 
where $H$ is the Hubble expansion rate.  Note that $K$ controls
whether the $N_1$ decays are in equilibrium, and is also a measure of
washout effects via inverse decays.
If we now take 
\begin{equation}
M_1~\sim 5\times 10^{12} \textrm{ GeV} 
\end{equation}
together with Eq.(\ref{eq:Mvalues}), we obtain 
$K \sim 0.3$.  Since $K\ll 1$ ($K \gg 1$) corresponds to the
weak (strong) washout regime, this parameter choice should lead to
moderate washout.
Moreover, for these parameters the light neutrino mass terms
become $m^A_\nu \sim 0.04$~eV and $m^B_\nu \sim 0.1$~eV, such that the
neutrino mass is dominated by the contribution from the diagram in
Fig.~\ref{fig:lightnumass}.  For a larger $M_1/\Lambda$ hierarchy,
$m^B_\nu$ would be suppressed with respect to $m^A_\nu$, but this is
undesirable as the asymmetry $\varepsilon$ would also be suppressed
by 4 powers of $M_1/\Lambda$. 

Finally, we note that effective dipole moment interactions of two
light neutrino states are induced by our Lagrangian.
The largest contributions arise from two-loop diagrams for which we
estimate
\begin{equation}
 \mu_{\nu_{\text{eff}}} \;\sim \left(\frac{\lambda'}{16\pi^2}\right)^2\,\frac{g'}{\Lambda}
 \;\simeq\; 5 \times 10^{-19}\; \mu_B\;,
\end{equation}
where the second approximate equality assumes the parameter values
specified above.  These induced dipole moments are thus well below
current experimental upper limits~\cite{numaglimit} which are of order
$10^{-11}\; \mu_B$.


\section{Conclusions}\label{sec:conclusion}


In summary, we have presented a new leptogenesis mechanism in which
electromagnetic dipole moment couplings induce $CP$ violating decays
of heavy RH neutrinos.  Via explicit calculation of the decay rates,
we have demonstrated that a sufficient asymmetry can be produced
through such decays to make this scenario viable.
However, since the electromagnetic dipole moment operators induce
neutrino mass terms, the leptogenesis parameters are constrained by the
masses of the usual light neutrinos.  For this reason, we find that
leptogenesis must take place at a high mass scale, comparable to that
for the standard scenario.
  
Washout effects, which reduce the asymmetry, will differ from the
standard scenario.  For example, the inverse decay is now a
$3\rightarrow 1$ process, and there will be a different set of $L$
violating scattering processes.  We leave the detailed study of these
washout effects to future work.
Finally, it is possible that the interference of the tree-level and
vertex amplitudes, which we have not explicitly calculated, may
enhance the asymmetry.  As with the standard mechanism, a near
degeneracy in the masses of the heavy neutrinos would also enhance the
self-energy contribution to the asymmetry~\cite{Covi:1996wh_Liu:1993tg,resonant}.

\section*{Acknowledgments}
We would like to thank Vincenzo Cirigliano, Bruce McKellar, Ray Volkas
and Lincoln Wolfenstein for helpful discussions.  NFB was supported by
the University of Melbourne Early Career Researcher Grant Scheme, BJK
by Fermilab (operated by the Fermi Resarch Alliance under US Dept. of
Energy contract DE-AC02-07CH11359) and SSCL by an Australian
Postgraduate Award.

\end{document}